\documentstyle[12pt,psfig]{article}
\newcommand{\EQ}{\begin{equation}}
\newcommand{\EN}{\end{equation}}
\newcommand{\bea}{\begin{eqnarray}}
\newcommand{\eea}{\end{eqnarray}}

\newcommand{\spz}{\hspace{0.7cm}}

\newcommand{\th}{\theta}
\newcommand{\var}{\varepsilon}

\begin{document}
\topmargin 0pt
\oddsidemargin 5mm
\renewcommand{\thefootnote}{\arabic{footnote}}
\newpage
\setcounter{page}{0}
\begin{titlepage}
\begin{flushright}
SISSA REF 16/2002/FM \\
DFTT 9/2002
\end{flushright}
\vspace{0.5cm}
\begin{center}
{\large {\bf On the fermion-boson correspondence for} \\
{\bf correlation functions of disorder operators}}\\
\vspace{1.8cm}
{\large G. Delfino$^a$, P. Grinza$^b$ and G. Mussardo$^{a}$} \\ 
\vspace{0.5cm}
{\em $^a$ SISSA and INFN, via Beirut 2-4, 34014 Trieste, Italy}\\
{\em $^b$ Dipartimento di Fisica and INFN,} \\
{\em Universit\`a degli Studi di Torino, via Giuria 1, 10125 Torino, Italy}
\end{center}
\vspace{1.2cm}

\renewcommand{\thefootnote}{\arabic{footnote}}
\setcounter{footnote}{0}

\begin{abstract}
\noindent
When a quantum field theory possesses topological excitations in a phase 
with spontaneously broken symmetry, these are created by operators which 
are non-local with respect to the order parameter. Due to non-locality, 
such disorder operators have non-trivial correlation functions even in free
massive theories. In two dimensions, these correlators can be expressed 
exactly in terms of solutions of non-linear differential equations. 
The correlation functions of the one-parameter family of non-local operators 
in the free charged bosonic and fermionic models are the inverse of each 
other. We point out a simple derivation of this correspondence within the 
form factor approach.
\end{abstract}

\end{titlepage}

\newpage
{\bf 1.}\,\,
The existence of excitations owing their stability to topological reasons
is a well known non-perturbative feature of quantum field theory. 
In a phase with spontaneously broken symmetry, field configurations may 
exist which interpolate at spatial infinity among different vacua of the 
theory and possess a topologic charge. In the quantum theory, these 
`extended' objects (solitons, vortices,..) must be created by the action
on the vacuum of operators which are non-local with respect to the 
order parameter. Since the operator content of the theory is expected 
to be the same on the two sides of the phase transition, 
such {\em disorder operators} can be identified 
also in the unbroken phase, where the topologic excitations are absent and 
they acquire non-zero vacuum expectation values.

In two space-time dimensions, only discrete symmetries can be broken 
spontaneously. Whenever this is the case, the broken phase possesses `kinks' 
interpolating between two different vacua. 
As a concrete example, consider the (euclidean) action
\EQ
{\cal A}=\int d^2x\,[\partial_\mu\phi^*\partial^\mu\phi+m^2\phi^*\phi
+W(\phi^*,\phi)]\,,
\label{action}
\EN
where the potential $W$ is invariant under the $Z_N$ transformation
$\phi\rightarrow e^{2i\pi/N}\phi$, $\phi^*\rightarrow e^{-2i\pi/N}\phi^*$. 
Denote $|0_j\rangle$, $j=0,1,\ldots,N-1$, the $N$ vacuua of the broken 
phase. Then we call $\mu_k(x)$, $k=1,\ldots,N-1$, the disorder operators 
which create the excitations interpolating between the vacua 
$|0_j\rangle$ and $|0_{j+k(mod\,N)}\rangle$. 
The operators $\mu_k(x)$ carry $k$ units of topological charge and 
satisfy the conjugation relation $\mu_k^*(x)=\mu_{N-k}(x)$. The mutual 
non-locality between the order and disorder operators manifests itself 
in a phase factor arising when they are taken around each other in the 
euclidean plane ($z=x_1+ix_2$, $\bar{z}=x_1-ix_2$)\,:
\bea
\phi(z e^{2i\pi},\bar{z}e^{-2i\pi})\,\mu_k(0,0) &=& e^{2i\pi k/N}
   \phi(z,\bar{z})\,\mu_k(0,0)\,,
\label{ac1}\\
\phi^*(z e^{2i\pi},\bar{z}e^{-2i\pi})\,\mu_k(0,0) &=& e^{-2i\pi k/N}
   \phi^*(z,\bar{z})\,\mu_k(0,0)\,.
\label{ac2}
\eea
These analytic continuations apply inside correlation functions and 
imply that the latter are not single valued. They hold along any 
$Z_N$-invariant renormalisation group trajectory flowing out of the 
phase transition point and characterise the operators $\mu_k(x)$ beyond 
their role of kink creation operators in the broken phase. 

The theory (\ref{action}) with $W=0$ is a particularly simple example 
of such a trajectory. Even in this case the correlation functions of 
disorder operators are non-trivial due to the non-locality with respect 
to the boson. One can hope, however, to compute them exactly exploiting 
the free particle basis, in analogy with what happens for the Ising field 
theory without magnetic field, where the spin correlators can be computed 
exactly due to the underlying free fermion theory \cite{McCoyWu}.

The correlation functions of non-local operators in two-dimensional free 
massive theories were extensively studied from the point of view of the 
deformation theory of differential equations in Ref.\,\cite{SMJ} and a 
series of related papers. Several works which include 
Refs.\cite{BKW,MSS,Alyosha,BB,BL} have been devoted afterwards to dealing
with this problem through more direct and general methods of quantum field
theory. In particular, the attention focused on the form factor approach
in which the correlators are expressed as sums over multiparticle asymptotic
states. While these spectral series can be written down explicitely for any
integrable quantum field theory, the free case is up to now the only one 
in which they can be resummed. The use of this approach to express the 
correlators through solutions of non-linear differential equations was 
illustrated in \cite{BB} and \cite{BL} for the neutral and charged fermionic
cases, respectively. 

In this note we point out that the form factor approach allows for a 
simple unified treatment of the bosonic and fermionic cases showing 
the correspondence\footnote{For a given operator $\Phi(x)$, the 
notation $\tilde{\Phi}(x)\equiv\Phi(x)/\langle\Phi\rangle$ will be used 
throughout this note.} \cite{SMJ}
\EQ
\langle\tilde{\mu}_j(x)\tilde{\mu}_k(0)\rangle
=\frac{1}{\langle\tilde{V}_{j/N}(x)\tilde{V}_{k/N}(0)\rangle}\,,
\label{main}
\EN
where the correlators on the l.h.s. are computed in the theory (\ref{action})
with $W=0$, and those on the r.h.s. refer to the operators
$V_\alpha(x)=\exp[i\sqrt{4\pi}\,\alpha\,\varphi(x)]$
in the sine-Gordon theory
\EQ
{\cal A}_{sG}=\int d^2x\left[\frac{1}{2}\,
\partial_\nu\varphi\partial^\nu\varphi-\mu\cos\beta\varphi\right]\,,
\label{sg}
\EN
with $\beta=\sqrt{4\pi}$ and $\mu$ a suitably chosen mass scale. 

We recall that the theory (\ref{sg}) with $\beta=\sqrt{4\pi}$ is in 
fact a free fermionic theory \cite{Coleman}. For generic values of $\beta$
the elementary excitations are the solitons and antisolitons interpolating 
between adjacent vacua of the periodic potential. Their interaction is 
attractive as long as $\beta<\sqrt{4\pi}$ and they form topologically 
neutral bound states, the lightest one being the particle interpolated by 
the bosonic field in the action (\ref{sg}). These neutral particles are 
absent from the spectrum of asymptotic states in the repulsive region
$\beta>\sqrt{4\pi}$, and at the point $\beta=\sqrt{4\pi}$ where the 
solitons behave as free Dirac fermions.

Since the solitons are non-local in terms of the field $\varphi(x)$,
the evaluation of both sides of Eq.\,(\ref{main}) amounts to computing
correlation functions of operators which are non-local with respect to 
non-interacting particles, bosons for the l.h.s. and fermions for the r.h.s. 
We will show how the different statistics produces the inversion in 
Eq.\,(\ref{main}). 

\vspace{.3cm}
{\bf 2.}\,\,
Consider a theory of free, charge conjugated particles $A$ and $\bar{A}$
with mass $m$, and denote by $\Phi_\alpha(x)$ a scalar operator exhibiting 
(in the sense of (\ref{ac1}), (\ref{ac2})) a non-locality phase 
$e^{2i\pi\alpha}$ ($e^{-2i\pi\alpha}$) with respect to (the field which 
interpolates) the particle $A$ ($\bar{A}$). Acting on the vacuum state 
$|0\rangle$, such an operator produces neutral states consisting of pairs 
$A(\theta)\bar{A}(\beta)$. We use rapidity variables to parameterise the 
energy-momentum of a particle as $(e,p)=(m\cosh\theta,m\sinh\theta)$. 
The equations satisfied by the form factors
\EQ
f^\alpha_n(\theta_1,\dots,\theta_n,\beta_1,\dots,\beta_n)=
\langle 0|\tilde{\Phi}_\alpha(0)|A(\theta_1),\dots,
A(\theta_n),\bar{A}(\beta_1),\dots,\bar{A}(\beta_n)\rangle
\label{ff}
\EN
are a particular case of those holding for generic integrable theories 
(see e.g.\,\cite{Alyosha})
\bea
f_n^\alpha(\theta_1,\dots ,\theta_i ,\theta_{i+1},\dots,\theta_n,\beta_1,
\dots,\beta_n) & = &
S\,f_n^\alpha(\theta_1,\dots,\theta_{i+1},\theta_i,\dots,\theta_n,\beta_1,
\dots,\beta_n), 
\label{ff1}\\
f_n^\alpha(\theta_1+2i\pi,\theta_2,\dots,\theta_n,\beta_1,\dots,\beta_n) &=& 
S\,e^{ 2 i \pi \alpha} 
f_n^\alpha(\theta_1, \dots,\theta_n,\beta_1, \dots,\beta_n), 
\label{ff2}\\
\textrm{Res}_{\theta_1-\beta_1= i \pi} f_n^\alpha(\theta_1, \dots,\theta_n,\beta_1, \dots,\beta_n)
& = & iS^{n-1}(1-e^{ 2 i \pi \alpha})f_{n-1}^\alpha(\theta_2,
..,\theta_n,\beta_2,..,\beta_n),
\label{ff3}
\eea
where 
\EQ
S=\left\{
\begin{array}{cl}
1 & \mbox{for free bosons} \\
-1 & \mbox{for free fermions}\,. \\
\end{array}
\right.
\EN
It is easy to check that the solution of the above system of 
equations is given by 
\EQ
f^\alpha_n(\theta_1,\dots,\theta_n,\beta_1,\dots,\beta_n)=
S^{n(n+2)/2}(-\sin\pi\alpha)^n\,e^{\left(\alpha-\frac12\delta_{S,1}\right)
\sum_{i=1}^n(\theta_i-\beta_i)}\,
\left|A_n\right|_{(S)}\,,
\label{solution}
\EN
where $A_n$ is a $n\times n$ matrix ($A_0\equiv 1$) with entries
\EQ
A_{ij}=\frac{1}{\cosh\frac{\theta_i-\beta_j}{2}}\,,
\EN
and $\left|A_n\right|_{(S)}$ denotes the 
permanent\footnote{The permanent of a matrix differs from the 
determinant by the omission of the alternating sign factors $(-1)^{i+j}$.} 
of $A_n$ for $S=1$ and the determinant of $A_n$ for $S=-1$. In fact, 
eq.\,(\ref{ff1}) (and the analogous equation for the permutation of 
two $\beta$ variables) immediately follows from the property of 
$\left|A\right|_{(S)}$ under exchange of two rows 
or columns. Concerning eq.\,(\ref{ff2}), the minus sign produced by 
$\left|A\right|_{(S)}$ when $\theta_1\rightarrow\theta_1 
+ 2i \pi$ is the one needed in the fermionic case, while it is canceled 
by the action of the Kroneker delta when $S=1$. Finally, eq.\,(\ref{ff3}) 
follows from the fact that taking the residue at $\theta_1-\beta_1=i\pi$ 
amounts to deleting the first row and the first column of $A_n$, which 
produces $\left|A_{n-1}\right|_{(S)}$.

The two-point correlators are given by
\EQ
G_{\alpha,\alpha'}^{(S)}(t)=
\langle\tilde{\Phi}_\alpha(x)\tilde{\Phi}_{\alpha'}(0)\rangle=
\sum_{n=0}^\infty\frac{1}{(n!)^2\,(2\pi)^{2n}}  
\int d\theta_1\dots d\theta_{n}d\beta_1\dots d\beta_{n} 
\,g_n^{(\alpha,\alpha')}(t\,|\theta_1,\dots,\beta_n)\,,
\label{corr}
\EN
where
\bea
g^{(\alpha,\alpha')}_n (t\,|\theta_1,\dots,\beta_n) &=& 
f_n^\alpha(\theta_1,\dots,\beta_n)f_n^{\alpha'}(\beta_n, \dots,\theta_1)\,
e^{-t e_n} \nonumber\\
&=& (S\sin\pi\alpha\,\sin\pi\alpha')^n\,
e^{(\alpha-\alpha')\sum_{i=1}^n(\theta_i-\beta_i)}\,
\left|A_n\right|_{(S)}^2 \,e^{-t e_n}\,,
\label{g}
\eea
\[
t=m|x|\,\,\,\,\,\,\,,
\,\,\,\,\,\,\,\,
e_n=\sum_{k=1}^n (\cosh\theta_k + \cosh\beta_k)\,\,\,.
\]
Analogously to the fermionic case \cite{BL}, we define
a new $n\times n$ matrix $M_n$ with entries
\EQ
M_{ij}\equiv M(\theta_i,\beta_j)=
(\sin\pi\alpha\,\sin\pi\alpha')^{1/2}\,
e^{-\frac{t}{2}\cosh\theta_i}\,
\frac{h(\theta_i)h^{-1}(\beta_j)}{\cosh\frac{\theta_i-\beta_j}{2}}\,
e^{-\frac{t}{2}\cosh\beta_j}\,\,\,,
\EN
\[
h(\theta)=e^{(\alpha-\alpha')\,\theta/2}\,,
\]
and rewrite eq.\,(\ref{g}) as
\EQ
g^{(\alpha,\alpha')}_n=S^n\left|M_n\right|^2_{(S)} =
\left|
\begin{array}{cc}
0 & M_n   \\
M_n^T  & 0 \\
\end{array}
\right|_{(S)}\,.
\EN
Finally we symmetrise with respect to the two sets of rapidities by 
introducing a charge index $\varepsilon$ which is $1$ for a particle $A$ 
and $-1$ for $\bar{A}$, so that we express eq.\,(\ref{corr}) as
\EQ
G_{\alpha,\alpha'}^{(S)}(t)=
\sum_{L=0}^\infty\frac{1}{L!\,(2\pi)^L}\sum_{\epsilon_1\dots\epsilon_L}\int 
d\theta_1\dots d\theta_{L}
\left|K_{\epsilon_i \epsilon_j}(\theta_i,\theta_j) 
\right|_{(S)}\,,
\label{fred}
\EN
where we have introduced the $L\times L$ matrices with entries
\bea
&& K_{+-}(\theta,\beta)=M(\theta,\beta)\,,\nonumber\\
&& K_{-+}(\theta,\beta)=M(\beta,\theta)\,, \label{pippo}\\
&& K_{++}(\theta,\beta)=K_{--}(\theta,\beta)=0\,\,\,.\nonumber
\eea
The last equation in (\ref{pippo}) ensures that only the terms with $L=2n$ 
occur in (\ref{fred}). Notice that the dependence on the statistics in 
(\ref{fred}) only reduces to taking the permanent or the determinant of 
the same matrix. According to the theory of Fredholm integral operators 
(see e.g.\,\cite{Schwinger}) this leads to the result
\EQ
G_{\alpha,\alpha'}^{(S)}(t)=\textrm{Det}(1+2\pi\,{\bf K})^{-S}\,,
\EN
so that the correlators of operators having the same non-locality 
phase in the free fermion and free boson theories are the inverse 
of each other. The result (\ref{main}) follows from the fact that the 
operators $V_\alpha(x)$ have a non-locality phase $e^{2i\pi\alpha}$ 
around the solitons at the sine-Gordon free fermion point (see 
e.g.\,\cite{af}), so that $\mu_j(x)$ and $V_{j/N}(x)$ have the same 
non-locality with respect to the corresponding particles. Since $j$ 
runs between $1$ and $N-1$, we are actually working with 
$0 <\alpha = j/N < 1$.

\vspace{.3cm}
{\bf 3.}\,\,
We have derived Eq.\,(\ref{main}) by using the large distance expansion 
of the correlators. It is interesting to check this inversion relation 
between the two correlators in the short distance limit. Both in the 
bosonic and fermionic case the operators $\Phi_\alpha(x)$ satisfy the 
operator product expansion
\EQ
\langle\Phi_\alpha(x)\Phi_{\alpha'}(0)\rangle\sim |x|^{-
\Gamma^{(S)}_{\alpha,\alpha'}}\langle\Phi_{\alpha+\alpha'}\rangle+\dots\,,
\label{ope}
\EN
with
\EQ
\Gamma^{(S)}_{\alpha,\alpha'}=X^{(S)}_{\alpha}+X^{(S)}_{\alpha'}-
X^{(S)}_{\alpha+\alpha'}\,\,\,,
\EN
$X^{(S)}_\alpha$ being the scaling dimensions. In the bosonic case the 
index $\alpha+\alpha'$ is taken modulo $1$.
The scaling dimensions can be computed through the formula 
\cite{DSC}
\EQ
X_\alpha^{(S)}= -\frac{1}{2\pi}\int d^2x
\langle\Theta(x)\tilde{\Phi}_\alpha(0)\rangle_{connected}\,,
\EN
where $\Theta(x)$ is the trace of the energy-momentum tensor. Since the only
non-zero form factor of this operator in the free theories is
\EQ
\langle 0|\Theta(0)|A(\theta)\bar{A}(\beta)\rangle=2\pi m^2\left[-i
\sinh\frac{\theta-\beta}{2}\right]^{\delta_{S,-1}}\,,
\EN
one easily finds
\EQ
X^{(S)}_\alpha=
\left\{
\begin{array}{ll}
\alpha(1-\alpha)\,\,\, , & S=1 \\
\alpha^2\,\,\, , & S=-1 \\
\end{array}
\right.
\EN
in agreement with the conjugation properties $\mu_j^*=\mu_{N-j}$ and 
$V_\alpha^*=V_{-\alpha}$. These results coincide with those of conformal 
field theory with `twist' fields (see \cite{Dixon,AlZam}). It follows
\EQ
\Gamma^{(-)}_{\alpha,\alpha'}=-2\alpha\alpha'\,\,\, ,
\EN
\EQ
\Gamma^{(+)}_{\alpha,\alpha'}=
\left\{
\begin{array}{ll}
2\alpha\alpha'\,, & \alpha+\alpha'<1 \\
2[\alpha\alpha'+1-(\alpha+\alpha')]\,, & \alpha+\alpha'>1\,\,\, . \\
\end{array}
\right.
\EN
The agreement for $1<\alpha+\alpha'<2$ is obtained by observing that in 
this range of $\alpha+\alpha'$, the leading short distance term in the 
fermionic case is not the one in (\ref{ope}) but the first off-critical 
contribution
\EQ
\mu/2\,\int d^2y\langle V_\alpha(x)V_\alpha(0)[V_1(y)+
V_{-1}(y)]\rangle_{\mu=0}\,\,\, ,
\label{off}
\EN
which indeed behaves as $|x|^{2[\alpha\alpha'+1-(\alpha+\alpha')]}$ when 
$x\rightarrow 0$.

In general, the two--point functions can be expressed as 
\EQ
G_{\alpha,\alpha'}^{(S)}(t)=e^{S\,\Upsilon_{\alpha,\alpha'}(t)}\,.
\label{newmain}
\EN
where $\Upsilon_{\alpha,\alpha'}(t)$ is given by 
\cite{SMJ,BL}
\EQ
\Upsilon_{\alpha,\alpha'}(t)=\frac12\int_{t/2}^\infty\rho d\rho\,
\left[(\partial_\rho\chi)^2-4\sinh^2\chi-
\frac{(\alpha-\alpha')^2}{\rho^2}\tanh\chi\right]\,,
\label{Sigma}
\EN
with $\chi(\rho)$ satisfying the differential equation
\EQ
\partial^2_\rho\chi+\frac{1}{\rho}\,\partial_\rho\chi=2\sinh 2\chi+
\frac{(\alpha-\alpha')^2}{\rho^2}\,\tanh\chi\,(1-\tanh^2\chi)\,.
\label{diff}
\EN
The precise short distance behaviour for $\alpha+\alpha'<1$ 
is
\EQ
\lim_{t\rightarrow 0}G_{\alpha,\alpha'}^{(S)}(t)=\left(C_{\alpha,\alpha'}\,
t^{2\alpha\alpha'}\right)^{-S}\,,
\label{uv}
\EN
with an amplitude that can be deduced from the work of Ref.\,\cite{LZ}
on vacuum expectation values in the sine-Gordon model:
\EQ
C_{\alpha,\alpha'}=2^{-2\alpha\alpha'}\exp\left\{2\int_0^\infty\frac{dt}{t}
\left[\frac{\sinh\alpha t\,\cosh(\alpha+\alpha')t\,\sinh\alpha't}{\sinh^2t}
-\alpha\alpha'e^{-2t}\right]
\right\}\,.
\label{ampl}
\EN

\vspace{.3cm}
{\bf 4.}\,\,
The case of $Z_2$ symmetry is somehow special. Since a broken phase with
two degenerate vacua can be realised in terms of a {\em neutral} boson, a 
disorder operator with non-locality factor $-1$ in present also in the theory 
of a neutral bosonic free particle. It is easy to check that, for $S=1$ and 
$\alpha=1/2$, the form factors (\ref{solution}) imply the factorisation
\EQ
\mu_1=\mu_{(1)}\times\mu_{(2)}\,,\hspace{1cm}N=2
\EN
where $\mu_{(j)}$ is the disorder operator with scaling dimension 
$X^{(+)}_{1/2}/2=1/8$ associated to the neutral boson $A_j$ entering the 
decomposition $A=(A_1+iA_2)/\sqrt{2}$, $\bar{A}=(A_1-iA_2)/\sqrt{2}$.

The fermion-boson correspondence observed above for charged particles has 
an analogue in the neutral case, and the correlation function
\EQ
G(t)=\langle\tilde{\mu}_{(j)}(x)\tilde{\mu}_{(j)}(0)\rangle=
\left[G^{(+)}_{1/2,1/2}(t)\right]^{1/2}
\label{neutral}
\EN
can be related to correlators computed in the theory of a free neutral 
fermion, i.e. in the 
Ising field theory without magnetic field. To see this, let us recall that
the correlators of the spin and disorder operators in the unbroken phase 
of the scaling Ising model can be written as \cite{McCoyWu}
\EQ
\tau_\pm(t)\equiv \langle\tilde{\mu}(x)\tilde{\mu}(0)\rangle\pm
                \langle\sigma(x)\sigma(0)\rangle=
\exp\left[\pm\frac12 \chi(t/2)-\frac12 \Upsilon_{1/2,1/2}(t)\right]\,,
\EN
where the functions $\chi$ and $\Upsilon$ are those of 
Eqs.\,(\ref{Sigma}), (\ref{diff}). Hence, it follows from (\ref{newmain}) 
and (\ref{neutral}) that
\EQ
G(t)=[\tau_+(t)\tau_-(t)]^{-1/2}\,.
\EN
Concerning the short distance behaviour of this correlator, the power law
(\ref{uv}) acquires in this case a logarithmic correction due to the 
`resonance' with the leading off-critical contribution (\ref{off}). Such 
a contribution to the correlator $\langle\tilde{V}_\alpha(x)
\tilde{V}_\alpha(0)\rangle$ behaves as $-C_{\alpha,\alpha}t^{1/2}
[1+(2-4\alpha)\ln t]$ in the limit 
$\alpha\rightarrow 1/2$, so that
\EQ
\lim_{t\rightarrow 0}G_{1/2,1/2}^{(-)}(t)=
\lim_{t\rightarrow 0}G^{-2}(t) = {\cal B}\,t^{1/2}\ln(1/t)\,,
\EN
with
\EQ
{\cal B} = - 4\,\mbox{Res}_{\alpha=1/2}\,C_{\alpha,\alpha}=0.588353..\,\,.
\EN
We checked that this amplitude coincides with that of the product 
$\tau_+(t)\tau_-(t)$ in the Ising model.

\vspace{.3cm}
{\bf 5.}\,\,
The free bosonic and fermionic theories discussed above can also be regarded 
as describing phases of spontaneously broken $Z_N$ symmetry. In this dual
vision, the excitations are free kinks $|K_{j,j\pm 1}(\theta)\rangle$ 
interpolating between two adjacent vacua $|0_j\rangle$ and 
$|0_{j\pm 1}\rangle$ (the indices are taken modulo $N$). We still denote by 
$\Phi_{k/N}$, $k=0,1,\dots,N-1$, the operators we are interested in. They 
correspond to the exponential operators $V_{k/N}$ in the fermionic 
case\footnote{In this case the N degenerate vacua can be identified with 
those of the periodic potential in (\ref{sg}) identified modulo N.}, and 
to the operators $\sigma_k$, dual to the disorder operators $\mu_k$, in 
the bosonic case. These operators create multikink excitations with 
zero topologic charge, i.e. starting and ending in the same vacuum state. 
Form factors on kink states were discussed in \cite{DC}. Consider 
for the sake of simplicity the two-kink matrix elements
\EQ
F_{j,k}^\pm(\theta_1-\theta_2)\equiv
\langle 0_j|\Phi_{k/N}(0)|K_{j,j\pm 1}(\theta_1)
K_{j\pm 1,j}(\theta_2)\rangle\,,
\label{ffkinks}
\EN
satisfying the equations 
\bea
F_{j,k}^\pm(\theta) & = & S\,F_{j,k}^\mp(-\theta)\,, 
\label{ffkink1} \nonumber \\
F_{j,k}^\pm(\theta+2i\pi) &=& F_{j\pm 1,k}^\mp(-\theta)\,, 
\label{ffkink2}\\
\textrm{Res}_{\theta=i\pi}F_{j,k}^\pm(\theta) & = & 
i\,\left[\langle 0_j|\Phi_{k/N}|0_j\rangle-\langle 0_{j\pm 
1}|\Phi_{k/N}|0_{j\pm 1}
\rangle \right]\,\,\, .
\label{ffkink3} \nonumber 
\eea
Since the generator $\Omega$ of $Z_N$ transformations ($\Omega^N=1$) acts 
on states and operators as ($\omega\equiv e^{2i\pi/N}$)
\bea
 \Omega\,|K_{j,j+1}(\theta_1)K_{j+1,j+2}(\theta_2)\dots\rangle &=&
           |K_{j+1,j+2}(\theta_1)K_{j+2,j+3}(\theta_2)\dots\rangle\,,
\nonumber \\
 \Omega^{-1}\,\Phi_{k/N}(x)\,\Omega & = &\omega^k\,\Phi_{k/N}(x)\,,
\nonumber 
\eea
the above form factor equations can be rewritten as 
\bea
F_{j,k}^\pm(\theta+2i\pi) &=& S\,\omega^{\pm k}F_{j,k}^\pm(\theta)\,,\\
\textrm{Res}_{\theta=i\pi}F_{j,k}^\pm(\theta) & = & 
i\,(1-\omega^{\pm k})\langle 0_j|\Phi_{k/N}|0_j\rangle\,.
\label{residue}
\eea
Once the identifications $A\longleftrightarrow 
K_{i,i+1}$, $\bar{A}\longleftrightarrow K_{i,i-1}$ are made,
these relations are equivalent to eqs.\,(\ref{ff2}), (\ref{ff3}) with 
$n=1$. The correspondence 
is easily extended to all values of $n$ and leads to the same form factors 
and correlation functions discussed for the unbroken phase. 

The introduction of a `magnetic field' pointing in the $k$ direction and
breaking explicitely the $Z_N$ symmetry corresponds to adding to the free 
action a term
\EQ
h\int d^2x\,\Psi_k(x)\,,
\EN
where 
\EQ
\Psi_k(x) = \frac{1}{N}\,\sum_{l=0}^{N-1}\omega^{-kl}\,
\Phi_{l/N}(x)\,\,\,. 
\EN
The first 
order corrections to the energy density $\varepsilon_j$ of the vacuum 
state $|0_j\rangle$ and to the mass $m_{j,j\pm 1}$ of the kink 
$K_{j,j\pm 1}$ are \cite{DMS}
\bea
&& \delta\varepsilon_j\sim h\,\langle 0_j|\Psi_k|0_j\rangle=h\,v\,\delta_{j,k}
\,,\\
&& \delta m_{j,j\pm 1}^2\sim h\,\langle 0_j|\Psi_k(0)|K_{j,j\pm 1}(i\pi)
K_{j\pm 1,j}(0)\rangle\,.
\eea
It follows from eq.\,(\ref{residue}) that 
\EQ
\textrm{Res}_{\theta=i\pi}\langle 0_j|\Psi_k(0)|K_{j,j\pm 1}(\theta)
K_{j\pm 1,j}(0)\rangle=\frac{iv}{N}\left(\delta_{j,k}-\delta_{j,k\mp 1}
\right)\,,
\EN
implying that the correction to the mass of the kinks interpolating 
between the vacua $|0_k\rangle$ and $|0_{k\pm 1}\rangle$ is 
infinite. This divergence simply reflects the fact that these kinks 
become unstable because the magnetic field removes the degeneracy of 
the vacuum $|0_k\rangle$ with the two adjacent vacua. These kinks 
are then confined in pairs $K_{k,k\pm 1}K_{k\pm 1,k}$ and the 
confinement gives rise to a string of bound states with zero 
topologic charge, as discussed in \cite{msg}.

\end{document}